\documentclass{IEEEtran}
\usepackage[utf8]{inputenc}
\DeclareUnicodeCharacter{03BC}{\textmu}
\usepackage[detect-all=true,exponent-product=\cdot]{siunitx}
\usepackage{algpseudocodex}

\usepackage{cite}

\ifCLASSINFOpdf
\else
\fi

\IfFileExists{builddir.def}{%
\input{builddir.def} 
}{%
\def\builddir{build}
}
\usepackage{tikz}
\usepackage[inkscapelatex=false,inkscapepath=\builddir/svg]{svg}
\usepackage{svg}

\usepackage[acronym]{glossaries}
\usepackage{xstring}

\newcommand\bit[1]{\num{#1} \unit{bit}}
\newcommand\bits[1]{\num{#1} \unit{bits}}
\newacronym{ai}{AI}{Artificial Intelligence}
\newacronym{conv}{CONV}{Convolutional}
\newacronym{dnn}{DNN}{Deep Neural Network}
\newacronym{fc}{FC}{Fully Connected}
\newacronym{ff}{FF}{Flip-Flop}
\newacronym{fi}{FI}{Fault Injection}
\newacronym{lut}{LUT}{Lookup Table}
\newacronym{mac}{MAC}{Multiply Accumulate}
\newacronym{nlf}{NLF}{Non Linear Function}
\newacronym{sa}{SA}{Systolic Array}
\newacronym{seu}{SEU}{Single-Event Upset}
\newacronym{ser}{SER}{Single-Event Upset Rate}
\newacronym{rtl}{RTL}{Register-Transfer Level}
\newacronym{relu}{ReLU}{Rectified Linear Unit}

\def\acrlist{}
\newcommand{\useacr}[1]{%
    \IfSubStr{\acrlist}{"#1"}{%
        \acrshort{#1}%
    }{%
        \expandafter\def\expandafter\acrlist\expandafter{\acrlist"#1"}%
        \acrlong{#1} (\acrshort{#1})%
    }%
}
\newcommand{\useacrplur}[1]{%
    \IfSubStr{\acrlist}{"#1"}{%
        \acrshort{#1}s%
    }{%
        \expandafter\def\expandafter\acrlist\expandafter{\acrlist"#1"}%
        \acrlong{#1}s (\acrshort{#1}s)%
    }%
}

\newcommand\SAsize[1]{#1$\times$#1}

\usepackage{float} 

\usepackage[font=footnotesize]{caption}
\usepackage[colorlinks=true,
            urlcolor=blue,
			citecolor=black,
			filecolor=black,
			linkcolor=black]{hyperref}

\begin{document}
%
\title{Single-Event Upset Analysis of a Systolic Array based Deep Neural Network Accelerator}

\author{%
    Naïn Jonckers\textsuperscript{1,2},
    Toon Vinck\textsuperscript{1,2,3},
    Gert Dekkers\textsuperscript{2},
    Peter Karsmakers\textsuperscript{3,4},
    Jeffrey Prinzie\textsuperscript{1}
    \\
    \textsuperscript{1}ESAT-ADVISE, KU Leuven, Belgium;
    \textsuperscript{2}Magics Technologies, Belgium;
    \textsuperscript{3}DTAI, Leuven.AI, KU Leuven, Belgium;
    \textsuperscript{4}MPRO, FLanders Make, Belgium
}
\maketitle

\begin{abstract}
\acrfull{dnn} accelerators are extensively used to improve the computational efficiency of
\acrshort{dnn}s, but are prone to faults through \acrlong{seu}s (\acrshort{seu}s).  In this work, we
present an in-depth analysis of the impact of \acrshort{seu}s on a \acrfull{sa} based \acrshort{dnn}
accelerator. A fault injection campaign is performed through a \acrfull{rtl} based simulation
environment to improve the observability of each hardware block, including the \acrshort{sa} itself
as well as the post-processing pipeline. From this analysis, we present the sensitivity, independent
of a \acrshort{dnn} model architecture, for various flip-flop groups both in terms of fault
propagation probability and fault magnitude. This allows us to draw detailed conclusions and
determine optimal mitigation strategies.
\end{abstract}



%
\IEEEpeerreviewmaketitle






\section{Introduction}
\label{sec:intro}

The advancements of \useacr{ai}, and specifically \useacrplur{dnn} in recent years,
has led to an increased interest in hardware acceleration to meet latency and energy requirements
driven by numerous applications. However, the majority of innovations have been geared towards
improving the performance of such accelerators, and not necessarily their reliability
\cite{bib:nvidia_error_prop}. In order to deploy this technology in safety-critical
applications, an assessment and potential improvement of its fault tolerance is required.

Concerning faults, an important contributor is the susceptibility of the hardware to \useacrplur{seu}.
Such a fault is typically caused by a high-energy particle, that can lead to a bit flip and possibly
even a system crash \cite{bib:dnn_safety}. To analyse the fault tolerance of an \useacr{ai}
accelerator, both radiation testing and simulations can be performed. While radiation tests are the
most realistic, they are time consuming and often provide less deep insights into how faults
propagate in the accelerator.

Prior work has reported on radiation effects in  \useacr{ai} accelerators. In
\cite{bib:libano_fpga}, neutron beam experiments have been conducted to obtain the model's
misclassification output cross-section. Similarly, a software-based \useacr{fi} analysis was
performed in \cite{bib:nvidia_error_prop}, which showed that the fault resilience depends on
data types, values, data reuse and the network layers. To avoid the limitations of \useacr{rtl}
simulation, the work in \cite{bib:sifiai} combines fast \useacr{ai} inference with \useacr{rtl}
simulation. Faults were only injected on the worst case registers in the data path of an \useacr{ai}
accelerator.

Existing literature is primarily focussed on the effects of faults at the output of \useacr{sa}
based accelerators, and not necessarily covers an in-depth sensitivity analysis of different
hardware blocks. Although the former is essential to assess reliability, it may offer less
information on how and where to include mitigation. Moreover, many works focus on the \useacr{sa}
and its accumulator \cite{bib:fi_google_tpu}, which may not give a full understanding on fault
propagation. Hence, this work will provide an in-depth analysis performed for different hardware
blocks of a \useacr{sa} based \useacr{dnn} accelerator by means of cycle-accurate \useacr{rtl}
simulations. This includes an analysis of the \useacr{sa} core itself as well as a post-processing
chain which, among others computes \useacrplur{nlf}. The analysis of this full pipeline (both
\useacr{sa} core and the post-processing chain) is essential to fully understand how faults
propagate.

At first the hardware architecture, including the full \useacr{sa} pipeline, is described in
section \ref{sec:arch}. Then the test methodology is presented in section \ref{sec:test},
elaborating on how the \useacr{fi} was carried out in simulation. The results of this analysis
are presented and discussed in section \ref{sec:result}. Finally, we present our conclusions
together with potential future work in section \ref{sec:concl}.

\section{Accelerator architecture}
\label{sec:arch}

\begin{figure}[H]
    \begin{center}
        \rotatebox{90}{%
        \resizebox{0.62\columnwidth}{!}{
{
\tikzstyle{arrow_sty}=[draw, very thick, ->, >=stealth]
\tikzstyle{op8bit_sty}=[draw, fill=blue!30!white, rounded corners=4pt]
\tikzstyle{op32bit_sty}=[draw, fill=red!30!white, rounded corners=4pt]
\tikzstyle{opmixed_sty}=[draw, fill=magenta!30!white, rounded corners=4pt]
\tikzstyle{input_sty}=[draw, fill=white!80!black]
\newcommand\bitsize[1]{\scriptsize(#1)}
\begin{tikzpicture}
    \node (sa) at (0,0)
          [opmixed_sty, minimum width=2cm, minimum height=2cm, text width=1.7cm, align=center]
          {SA\\\bitsize{8$\rightarrow$32 bit}};
    \node (acc) at ([yshift=-1cm]sa.south)
          [op32bit_sty, minimum width=2cm, minimum height=1cm, text width=1.7cm, align=center]
          {Accumulate\\\bitsize{32 bit}};
    \node (round) at ([yshift=-1cm]acc.south)
          [opmixed_sty, minimum width=2cm, minimum height=1cm, text width=1.7cm, align=center]
          {Round\\\bitsize{32$\rightarrow$8 bit}};
    \node (nlf) at ([yshift=-1cm]round.south)
          [op8bit_sty, minimum width=2cm, minimum height=1cm, text width=1.7cm, align=center]
          {NLF\\\bitsize{8 bit}};
    \node (inp_a) at ([xshift=-1cm]sa.west)
          [input_sty, minimum width=2cm, minimum height=1cm, rotate=-90, text width=1.7cm, align=center]
          {Activations\\\bitsize{8 bit}};
    \node (inp_w) at ([yshift=1cm]sa.north)
          [input_sty, minimum width=2cm, minimum height=1cm, text width=1.7cm, align=center]
          {Weights\\\bitsize{8 bit}};
    \node (inp_b) at ([xshift=-1.5cm]acc.west)
          [input_sty, minimum width=2cm, minimum height=1cm, text width=1.7cm, align=center]
          {Biases\\\bitsize{32 bit}};
    \node (inp_sr) at ([xshift=-1.5cm]round.west)
          [input_sty, minimum width=2cm, minimum height=1cm, text width=1.7cm, align=center]
          {Right shift\\\bitsize{5 bit}};
    \node (inp_nlf) at ([xshift=-1.5cm]nlf.west)
          [input_sty, minimum width=2cm, minimum height=1cm, text width=1.7cm, align=center]
          {Non-linear function};
    \node (outp) at ([yshift=-1cm]nlf.south)
          [input_sty, minimum width=2cm, minimum height=1cm, text width=1.7cm, align=center]
          {Outputs\\\bitsize{8 bit}};

    \node (gemm_group) at ([xshift=0.75cm]sa.south east) []{\resizebox{0.7cm}{3cm}{\}}};
    \node (gemm_name) at (gemm_group.east) [rotate=-90]{Matrix multiplication};
    \node (postproc_group) at ([xshift=0.75cm,yshift=0.75cm]nlf.east) []{\resizebox{0.7cm}{2cm}{\}}};
    \node (postproc_name) at (postproc_group.east) [rotate=-90]{Post-processing};

    \draw[arrow_sty] (inp_a.north)      -- (sa.west);
    \draw[arrow_sty] (inp_w.south)      -- (sa.north);
    \draw[arrow_sty] (inp_b.east)       -- (acc.west);
    \draw[arrow_sty] (inp_sr.east)      -- (round.west);
    \draw[arrow_sty] (inp_nlf.east)     -- (nlf.west);
    \draw[arrow_sty] (nlf.south) -- (outp.north);

    \draw[arrow_sty] (sa.south)                    -- (acc.north);
    \draw[arrow_sty] ([shift={(0,-0.2)}]acc.south) -- ([shift={(1.3,-0.2)}]acc.south) |- (acc.east);
    \draw[arrow_sty] (acc.south)                   -- (round.north);
    \draw[arrow_sty] (round.south)                 -- (nlf.north);
    \draw[arrow_sty] (nlf.south)                   -- (outp.north);
\end{tikzpicture}
}}
        }
    \end{center}
    \vspace{-1.7em}
    \caption{Block diagram of the accelerator pipeline}
    \label{fig:accelerator_pipeline}
\end{figure}
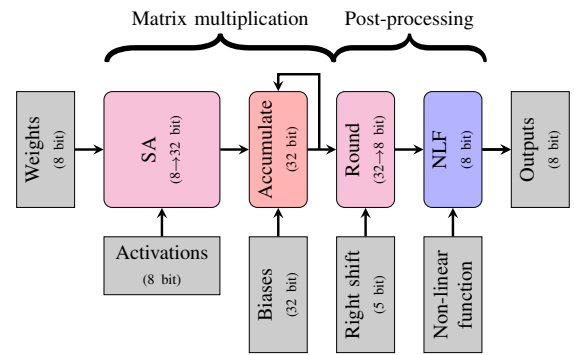

The research, described in this paper, was performed on a \useacr{sa} based \useacr{dnn}
accelerator. In addition to the \useacr{sa}, we also implemented a post-processing pipeline
containing accumulation rounding and \useacr{nlf} activation. \bit{8} input activations $a_r^{i8}$
and weights $w_{r\times c}^{i8}$ are sent to the \useacr{sa}, which then accumulates partial matrix
multiplication results in a \bit{32} register together with a \bit{32} bias. Afterwards, in the
post-processing pipeline, the \bit{32} result is scaled back to a final \bit{8} result and sent
through the \useacr{nlf} block. We define a single propagation of data from input to output as an
\textit{iteration} of the accelerator pipeline.  Figure \ref{fig:accelerator_pipeline} shows the
conceptual block diagram of the entire accelerator pipeline.

\vspace{-1em}

\begin{equation}\label{eqn:gemm}
    y_c^{i32} = b_c^{i32} + \sum_{r=1}^{R} \left( a_r^{i8} w_{r\times c}^{i8} \right)
\end{equation}

\useacr{fc} or ``dense'' layers of a \useacr{dnn}, which are essentially matrix multiplications, are
computed by the \useacr{sa} of the accelerator. If the dimensions of a \useacr{fc} layer are larger
than the \useacr{sa} dimensions, partial results are accumulated in the accumulator. The amount of
rows and columns of the accelerator is parametrised to facilitate the analysis in section
\ref{sec:test}. \useacr{conv} layers are first transformed into matrix multiplications by the
instruction compiler, after which they are computed by the \useacr{sa}. Figure \ref{fig:sa_detail}
highlights the data flow inside a \useacr{sa} with 2 rows and 2 columns where the matrix
multiplication is computed by means of several \useacr{mac} operations. Since the implemented
\useacr{sa} is weight-stationary, weights are preloaded into a register per \useacr{mac} element.
Activations are skewed by delaying them and streamed from bottom to top while partial sums
($a_r^{i8} w_{r\times c}^{i8}$) flow from left to right and are skewed again at the output so that they arrive
simultaneously at the accumulator.  This accumulator, which is implemented for every \useacr{sa}
column, is preloaded with the bias $b_c^{i32}$ for that particular column and thus after one or multiple
accumulation cycles the full matrix multiplication, shown in equation \ref{eqn:gemm}, is computed
\cite{bib:whitepaper_quantization}.  Here, the suffix $c$ stands for ``column'' and $r$ stands for
``row'' corresponding to rows and columns of the \useacr{sa} pipeline.

\vspace{-1.2em}
\begin{figure}[H]
    \begin{center}
        \rotatebox{90}{%
        \resizebox{0.8\columnwidth}{!}{
{
\tikzstyle{arrow_sty}=[draw, very thick, ->, >=stealth]
\tikzstyle{arrowdashed_sty}=[draw, dashed, thick, ->, >=stealth]
\tikzstyle{reg8_sty}=[draw, fill=blue!20!white]
\tikzstyle{reg32_sty}=[draw, fill=red!20!white]
\tikzstyle{mac_sty}=[draw, fill=white]
\tikzstyle{input_sty}=[draw, fill=white!80!black]
\newcommand\DrawHReg[2]{%
    \node (#1) at (#2) [reg8_sty, minimum height=2cm, minimum width=0.5cm]{};
    \node (#1_text) at (#2) [rotate=-90]{\scriptsize 8-bit reg};
    \draw ([xshift=0.1cm]#1.north) -- ([yshift=-0.3cm]#1.north) -- ([xshift=-0.1cm]#1.north);
}
\newcommand\DrawVReg[2]{%
    \node (#1) at (#2) [reg32_sty, minimum height=0.5cm, minimum width=2cm]{};
    \node (#1_text) at (#2) []{\scriptsize 32-bit reg};
    \draw ([yshift=0.1cm]#1.west) -- ([xshift=0.3cm]#1.west) -- ([yshift=-0.1cm]#1.west);
}
\newcommand\DrawWReg[2]{%
    \node (#1) at (#2) [reg8_sty, minimum height=0.5cm, minimum width=5cm]{};
    \node (#1_text) at (#2) []{\scriptsize \SAsize{2} 8-bit reg};
    \draw ([yshift=0.1cm]#1.west) -- ([xshift=0.3cm]#1.west) -- ([yshift=-0.1cm]#1.west);
}
\newcommand\DrawMac[3]{%
    \node (#1) at (#2)
          [mac_sty, minimum width=2cm, minimum height=2cm, text width=1.7cm, align=center]{};
    \node (#1_text) at (#1) [rotate=-90, text width=2cm, align=center]{MAC\\#3};
}
\begin{tikzpicture}
    \DrawWReg{wreg}{1.5,5}
    \foreach \row in {0,...,1}{
        \foreach \col in {0,...,1}{
            \pgfmathsetmacro\realrow{int(1-\row)}
            \node (mac_r\realrow c\col_dummy) at (3*\col, 3*\row)
                  [minimum width=2cm, minimum height=2cm]{};
        }
    }
    \draw[arrowdashed_sty] ([xshift=-2cm]wreg.south) -- ([xshift=-0.5cm]mac_r0c0_dummy.north);
    \draw[arrowdashed_sty] ([xshift=-2cm]wreg.south) -- ([xshift=-0.5cm]mac_r1c0_dummy.north);
    \draw[arrowdashed_sty] ([xshift=1cm]wreg.south) -- ([xshift=-0.5cm]mac_r0c1_dummy.north);
    \draw[arrowdashed_sty] ([xshift=1cm]wreg.south) -- ([xshift=-0.5cm]mac_r1c1_dummy.north);
    \foreach \row in {0,...,1}{
        \foreach \col in {0,...,1}{
            \pgfmathsetmacro\realrow{int(1-\row)}
            \DrawMac{mac_r\realrow c\col}{3*\col, 3*\row}{$r=\realrow$\\$c=\col$}
        }
    }
    \DrawHReg{ffchain_h_r0i1}{-2,3}
    \DrawHReg{ffchain_h_r1i1}{-2,0}
    \DrawHReg{ffchain_h_r1i0}{-3,0}
    \DrawVReg{ffchain_v_c0i0}{0,-2}
    \DrawVReg{ffchain_v_c0i1}{0,-3}
    \DrawVReg{ffchain_v_c1i0}{3,-2}
    \DrawHReg{sa_hreg_r0c0}{1.5,3}
    \DrawHReg{sa_hreg_r1c0}{1.5,0}
    \DrawVReg{sa_vreg_r0c0}{0,1.5}
    \DrawVReg{sa_vreg_r0c1}{3,1.5}
    \node (inp_a) at (-4.25,1.5)
          [input_sty, minimum width=5cm, minimum height=1cm, rotate=-90]
          {Activations $[a_1, a_0]$};
    \node (inp_w) at (1.5,6.25)
          [input_sty, minimum width=5cm, minimum height=1cm]
          {\footnotesize Weights $[[w_{0\times0}, w_{0\times1}], [w_{1\times0}, w_{1\times1}]]$};
    \node (outp) at (1.5,-4.25)
          [input_sty, minimum width=5cm, minimum height=1cm]
          {Outputs $[y'_0, y'_1]$};

    \draw[arrow_sty] (ffchain_h_r0i1.east) -- (mac_r0c0.west);
    \draw[arrow_sty] (mac_r0c0.east)       -- (sa_hreg_r0c0.west);
    \draw[arrow_sty] (sa_hreg_r0c0.east)   -- (mac_r0c1.west);
    \draw[arrow_sty] (ffchain_h_r1i0.east) -- (ffchain_h_r1i1.west);
    \draw[arrow_sty] (ffchain_h_r1i1.east) -- (mac_r1c0.west);
    \draw[arrow_sty] (mac_r1c0.east)       -- (sa_hreg_r1c0.west);
    \draw[arrow_sty] (sa_hreg_r1c0.east)   -- (mac_r1c1.west);
    \draw[arrow_sty] (mac_r0c0.south)       -- (sa_vreg_r0c0.north);
    \draw[arrow_sty] (sa_vreg_r0c0.south)   -- (mac_r1c0.north);
    \draw[arrow_sty] (mac_r1c0.south)       -- (ffchain_v_c0i0.north);
    \draw[arrow_sty] (ffchain_v_c0i0.south) -- (ffchain_v_c0i1.north);
    \draw[arrow_sty] (mac_r0c1.south)       -- (sa_vreg_r0c1.north);
    \draw[arrow_sty] (sa_vreg_r0c1.south)   -- (mac_r1c1.north);
    \draw[arrow_sty] (mac_r1c1.south)       -- (ffchain_v_c1i0.north);
    \draw[arrow_sty] (inp_w.south) -- (wreg.north);
    \draw[arrow_sty] ([yshift=1.5cm]inp_a.north)  -- (ffchain_h_r0i1.west);
    \draw[arrow_sty] ([yshift=-1.5cm]inp_a.north) -- (ffchain_h_r1i0.west);
    \draw[arrow_sty] (ffchain_v_c0i1.south) -- ([xshift=-1.5cm]outp.north);
    \draw[arrow_sty] (ffchain_v_c1i0.south) -- ([xshift=1.5cm]outp.north);

    \draw[arrow_sty, line width=0.1cm, red]
        ([xshift=0.25cm,yshift=0.05cm]mac_r0c1.north east) --
        ([xshift=0.25cm,yshift=-0.5cm]mac_r1c1.south east)
        node[anchor=west, text width=2.5cm, rotate=-90]{partial sum flow};
    \draw[arrow_sty, line width=0.1cm, blue]
        ([xshift=-0.5cm,yshift=0.25cm]mac_r0c0.north west)
        node[anchor=east, text width=1.5cm, rotate=-90, align=right, yshift=-0.2cm]{activation flow}
        -- ([xshift=0.5cm,yshift=0.25cm]mac_r0c1.north east);
\end{tikzpicture}
}}
        }
    \end{center}
    \vspace{-2em}
    \newbox\boxFigSA
    \begin{lrbox}{\boxFigSA}
        \footnotesize Detailed view of a \SAsize{2} \useacr{sa}
    \end{lrbox}
    \caption{\usebox{\boxFigSA}}
    \label{fig:sa_detail}
\end{figure}

After accumulating the output of the \useacr{sa}, we may end up with a result that is larger than
\bits{8} which, in order to save it back to memory, needs to be re-quantised to \bits{8}. In
\useacr{ai} frameworks like PyTorch, these calculations are typically performed in \bit{32} IEEE
floating-point format. In order to mimic integer-only arithmetic, the values are typically
\textit{fake-quantised} in-between operations \cite{bib:whitepaper_quantization}. Equation
\ref{eqn:quant} applies symmetric quantisation to a floating-point number $x^{f32}$ to obtain a
fake-quantised floating-point number $\hat{x}^{f32}$, with $s$ a floating-point scale factor
(smaller than 1) and $x_{i8}$ an \bit{8} integer number.
\begin{equation}\label{eqn:quant}
    \hat{x}^{f32} = \mathrm{clip}\big(\mathrm{round}(\tfrac{x^{f32}}{s})\big)\cdot s = x^{i8}\cdot s
\end{equation}

Combining equation \ref{eqn:gemm} and \ref{eqn:quant} leads to equation \ref{eqn:gemm_quant}, with
$s_a$, $s_w$ and $s_y$ the scale factor for the input activation, weight and output respectively
\cite{bib:whitepaper_quantization}.
\begin{equation}\label{eqn:gemm_quant}
    y^{i32}_c = \mathrm{clip}\left(\mathrm{round}\left(
        \left(b^{i32}_c + \sum_{r=1}^{R} \left( a^{i8}_r w^{i8}_{r\times c} \right)\right)
         \frac{s_a  s_w}{s_y}
    \right)\right)
\end{equation}

If we implement the scaling operation in the rounding block of the accelerator, the critical path of
that block would become quite large because of the division required for the scaling factors.
Therefore, we constrain $s_a$, $s_w$ and $s_y$ to a power of two, after which we can use a simple
right shift ($S$) operation.
\begin{equation}\label{eqn:scale_to_shift}
    s = \frac{s_a  s_w}{s_y} \quad\stackrel{s \in 2^n}{\Rightarrow}\quad
    S = \log_2\left(\frac{1}{s}\right)
\end{equation}
\begin{equation}\label{eqn:gemm_quant_shift}
    y^{i32}_c = \mathrm{clip}\left(\mathrm{round}\left(
        \left(b^{i32}_c + \sum_{r=1}^{R} \left( a^{i8}_r w^{i8}_{r\times c} \right)\right) >> S
    \right)\right)
\end{equation}

Aside from matrix multiplications in \useacr{fc} and \useacr{conv} layers and the rounding of the
accumulated \bit{32} result, \useacrplur{dnn} also use non-linear functions such as \useacr{relu} to model
non-linear effects. These functions are implemented in the accelerator pipeline in the \useacr{nlf}
block, which is implemented as a \useacr{lut} after which the result can be written back to memory.

\section{Test methodology}
\label{sec:test}

\subsection{Experiment setup}
\label{sec:test:setup}

The goal of our experiment is to analyse the sensitivity of different components in the accelerator
pipeline to \useacrplur{seu}. Our setup uses a simulation-based \useacr{rtl} fault injection
methodology where a single \useacr{seu} is applied to a random \useacr{ff} at a random cycle
during a single iteration.

In real environments, the accelerator can be hit by multiple \useacrplur{seu} in a single iteration
(i.e. \useacr{dnn} inference). However, the probability of the accelerator being hit by more than
one \useacr{seu} in the same iteration is very low. Wang et al., measured the cross section of a
\useacr{ff} in 28-nm bulk technology \cite{bib:seu_28nm_fdsoi}. Using the OMERE simulation tool, we
estimated a \useacr{ser} of $2.82 \ \cdot 10^{-7}$ \unit{SEU/FF/day} in GEO orbit, with an aluminium
shield of 100 \unit{mil}. Using this \useacr{ser}, the probability of the accelerator being hit by
$k$ \useacrplur{seu} in a single iteration can be estimated using the Poisson distribution
\cite{bib:pitfalls_fault_injection}:

\begin{equation}\label{eqn:poisson}
    P(k \ \text{SEUs}) = \frac{\lambda^k e^{-\lambda}}{k!}
\end{equation}

In this work, $\lambda$ is the expected number of \useacrplur{seu} during one clock cycle. As
indicated in equation \ref{eqn:fault_lambda}, this factor can be calculated as the \useacr{ser} multiplied by the number of \useacrplur{ff} ($N_{ff}$) in the accelerator pipeline
divided by the clock frequency ($f_{clock}$) of this accelerator. Table \ref{tab:seu_prob} shows
that, even for a relatively low frequency of \num{100} \unit{MHz}, the probability of multiple
upsets ($k > 1$) is negligible. An increase in the size of the \useacr{sa} has some impact, but also
there, the probability is still low enough to justify the use of a single \useacr{fi} for one
iteration of data through the accelerator pipeline.

\begin{equation}\label{eqn:fault_lambda}
    \lambda = \frac{N_{ff} \text{\useacr{ser}}}{f_{clock}}
\end{equation}

\vspace{-0.5em}
\begin{table}[h!]
    \newbox\boxTabProb
    \begin{lrbox}{\boxTabProb}
        \footnotesize \useacrplur{seu}
    \end{lrbox}
    \caption{
        Estimated probability of $k$ \usebox{\boxTabProb} in a \SAsize{2}, \SAsize{4} and \SAsize{8}
        accelerator in GEO orbit for a clock frequency of \num{100} \unit{MHz}
    }
    \label{tab:seu_prob}
    \begin{center}
        \begin{tabular}{|c|c c c|}
            \hline
                $k$ &
                $P_{\lambda2\times2}(k \ \text{\useacrplur{seu}})$ &
                $P_{\lambda4\times4}(k \ \text{\useacrplur{seu}})$ &
                $P_{\lambda8\times8}(k \ \text{\useacrplur{seu}})$
            \\ \hline
                $n_{ffs}$ & \num{352} & \num{1240} & \num{4648}
            \\ \hline
                0 & \num{0.999999}... & \num{0.999999}... & \num{0.999999}... \\
                1 & \num{1.149e-17} & \num{4.047e-17} & \num{1.517e-16} \\
                2 & \num{6.600e-35} & \num{8.190e-34} & \num{1.151e-32} \\
            \hline
        \end{tabular}
    \end{center}
\end{table}
\vspace{-1.3em}



\subsection{Single-Event Upset injection testbench}
\label{sec:test:seu}

Given the constraint of a single fault per iteration, we implement a fault injection testbench that
injects uniform random faults into the accelerator's \useacrplur{ff}.
The list of all \useacrplur{ff} is extracted from the elaborated netlist of the RTL design. This
testbench then wraps around the functional verification testbench in order to efficiently
observe functional failures (i.e. faults that propagate towards the output) as faults are injected.
This \useacr{seu} injection methodology is then repeated for multiple iterations of data and
multiple variations of the \useacr{sa} size, specifically \SAsize{2}, \SAsize{4} and \SAsize{8}.
Additionally, we also constrain the functional testbench inputs as described below.

\subsection{Constraining inputs}
\label{sec:test:random}

The purpose of the \useacr{seu} testbench is to evaluate the sensitivity to faults of each component
in the accelerator pipeline in a way that mimics the behaviour of a representative \useacr{dnn}
workload, though independent of a specific \useacr{dnn} model. This is done by generating
constrained random data, sampled from distributions that represent typical \useacr{dnn} workloads.

As visualised in figure \ref{fig:accelerator_pipeline}, the accelerator pipeline consists of four
different inputs: weights, activations, biases and a right shift value.  We assume that the weights
and activations in a \useacr{dnn} are approximately normally distributed.
As this accelerator is designed to work with quantized numbers, the floating point numbers have to
be scaled, rounded and clipped to the range of the quantized numbers. Therefore, a good choice for
scaling factors is required to mimic the behaviour of a \useacr{dnn} workload.  For the weights and
activations we use a normal distribution $\mathcal{N}(0, 1)$. The activation, weight and output
scale factor are determined by a $3\sigma$ calibration, rounded up to a power of two
\cite{bib:whitepaper_quantization}.

The distribution of the biases is less understood, therefore we use the standard PyTorch
initialisation for biases in linear layers. This distribution is defined as
$\mathcal{U}\left(-\frac{1}{\sqrt{j}}, \frac{1}{\sqrt{j}}\right)$, with $j$ the row size. It is
scaled with the product of the weight and activation scale factor
\cite{bib:whitepaper_quantization}.

As indicated by formula \ref{eqn:scale_to_shift} and \ref{eqn:gemm_quant_shift} the right shift
value is a fixed value that is chosen based on the quantization settings of the weights and
activations. \useacr{relu} is chosen as the non-linear function in this experiment.

Using these constrained random inputs, with the \useacr{fi} strategy as described in section
\ref{sec:test:seu}, we ran 10 million iterations of the testbench. This causes the confidence
interval in the results plots to be small enough so that this can be omitted. The results of this
experiment are presented in section \ref{sec:result}.

\section{Results}
\label{sec:result}

In this section, we will apply a comparative analysis of the sensitivity to bit faults of various
components within the accelerator pipeline. For the purposes of this analysis, we divide the
accelerator pipeline's numerous registers into eight separate groups according to their diversity:
\begin{enumerate}
    \item \textit{sa-reg-ffchain-h}: Horizontal activation input register chain before \useacr{mac}
          units of \useacr{sa}
    \item \textit{w-reg}: Weight input registers before \useacr{mac} units
    \item \textit{sa-reg-h}: Horizontal registers inside \useacr{sa}
    \item \textit{sa-reg-v}: Vertical registers inside \useacr{sa}
    \item \textit{sa-reg-ffchain-v}: Vertical output register chain after \useacr{mac} units of
          \useacr{sa}
    \item \textit{accum-reg}: Accumulator register for each column
    \item \textit{round-reg}: Register per column for each round block
    \item \textit{nlf-reg}: Register per column for each \useacr{nlf} \useacr{lut}
\end{enumerate}
We also split the sensitivity into two distinct analysis parts:
\begin{enumerate}
    \item \textbf{Fault Propagation Probability}: The probability of a fault at the output of the
          accelerator pipeline, given a fault in a specific component.
    \item \textbf{Fault Magnitude}: The absolute magnitude of the fault at the output of the
          accelerator pipeline, given a fault in a specific component.
\end{enumerate}

\begin{figure}[!ht]
\vspace{-1em}
    \begin{center}
        \includegraphics[scale=0.4]{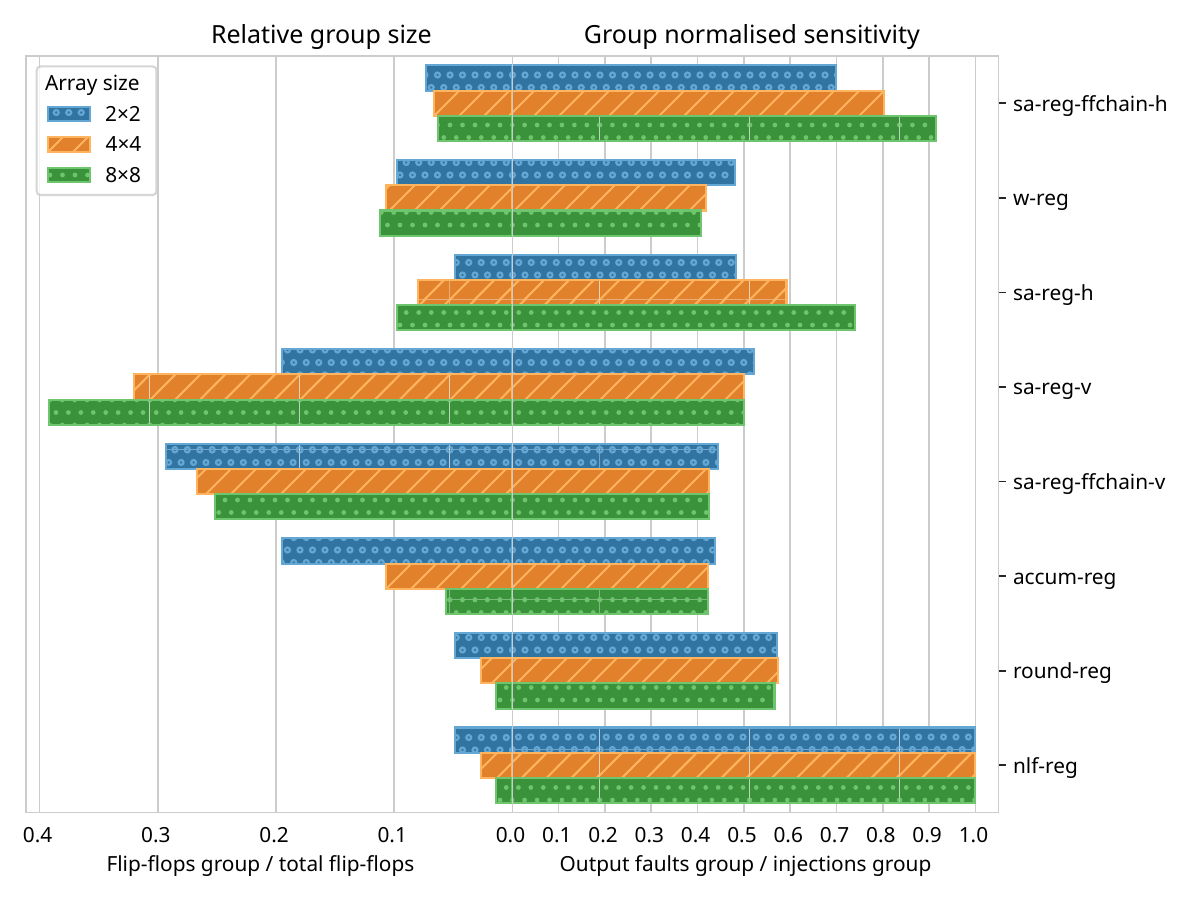}
    \end{center}
    \newbox\boxFigResult
    \begin{lrbox}{\boxFigResult}
        \footnotesize\useacrplur{ff}
    \end{lrbox}
    \caption{
        \textbf{Left}: The ratio of the number of \usebox{\boxFigResult} in that group to the total
                       number of \usebox{\boxFigResult} in the accelerator;
        \textbf{Right}: The ratio of the number of faults at the output to the number of injections
                        in that group
    }
    \label{fig:combined_sensitivity_contribution}
    \vspace{-1em}
\end{figure}

Figure \ref{fig:combined_sensitivity_contribution} shows the fault propagation probability of
each group in the accelerator pipeline. On the right side it shows the groups' normalised
sensitivity, which is calculated as the ratio of the number of faults at the output to the number of
injections in that group.

Firstly, it is important to indicate that, not every bit fault in the accelerator generates an
output fault. This is due to the masking effect of the rounding block, which can shift the fault
below the least significant bit. Secondly, the \useacr{relu} \useacr{nlf} also masks all negative values to
zero. An exception to this rule are the registers in the \textit{nlf-reg} group, which are
positioned after these masking operations. Therefore, they do not experience this masking effect.
Since the \textit{round-reg} group is positioned after the rounding operation, but before the
\useacr{nlf}, it only benefits from the masking effect of the \useacr{relu}. All other groups experience
both masking effects, which reduces the probability of a fault propagating to the output.

The \textit{sa-reg-ffchain-h} and \textit{sa-reg-h} groups become more sensitive as the \useacr{sa}
size grows since a fault, injected in these groups can propagate to multiple columns. This increases
the chance of a positive value that is not masked by the \useacr{relu}.

The left side of figure \ref{fig:combined_sensitivity_contribution} shows the number of
\useacrplur{ff} in each group, relative to the entire accelerator pipeline. As stated by Amdahl's
law, in order to achieve the maximal improvement gained from hardening certain groups, it is
important to not only focus on the most sensitive components, but also on the amount of
\useacrplur{ff} present in these groups. To ensure a maximal overall improvement, both sides of
figure \ref{fig:combined_sensitivity_contribution} must be taken into account.

\begin{figure}[!th]
\vspace{-1em}
    \begin{center}
        \includegraphics[scale=0.45]{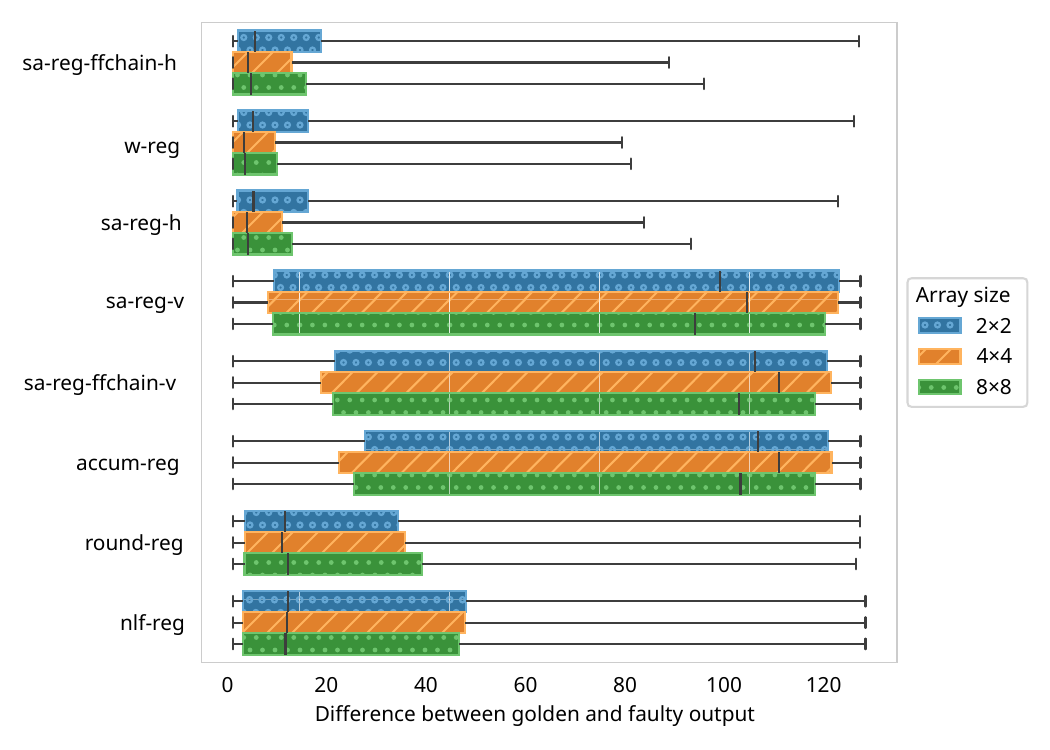}
    \end{center}
    \caption{Output fault magnitude distribution for each group}
    \label{fig:box_fault_size}
    \vspace{-1em}
\end{figure}

Figure \ref{fig:box_fault_size} shows a representation of the fault magnitude. The box-plots in this
figure visualises the distribution of the maximum fault magnitude in each output vector element,
when injected in a certain group. The whiskers in this box-plots represent the minimum and maximum
value in this distribution. From these results, there is no clear trend between the different
accelerator sizes. However, different groups clearly differentiate in fault magnitude. A distinction
between three collections of groups can be made. The first collection are the
\textit{sa-reg-ffchain-h}, \textit{w-reg} and \textit{sa-reg-h} groups. These groups consist of
\bit{8} registers and they are located before the accumulator. They have predominantly small output
fault sizes, which can be attributed to the fact that faults after the MAC operations are scaled
down by the right shift operation. The second collection are the \bit{32} registers, which are the
\textit{sa-reg-v}, \textit{sa-reg-ffchain-v} and \textit{accum-reg} groups. These groups are also
scaled down by the right shift operation, but clearly show larger output fault magnitudes. This can
be explained because \bit{32} registers have a higher dynamic range, making them more prone to
faults. The last group contains the \bit{8} registers after the accumulator, which are the
\textit{round-reg} and \textit{nlf-reg}. Faults occurring in the \textit{nlf-reg} are directly fed
to the output, resulting in a fault size of a power of two. Faults in \textit{round-reg} differ
from \textit{nlf-reg}, since they can be masked by the \useacr{relu} operation. This results in a slightly
smaller fault magnitude distribution.

The results from fault propagation probability analysis show that the post-processing chain
registers \textit{nlf-reg} and \textit{round-reg} are prone to fault propagation. However, the
amount of \useacrplur{ff} in these groups is relatively small and their average fault magnitude is
also small. This implies that the gain in hardening these registers is also relatively small.

Secondly, our analysis shows that the \bit{8} register groups \textit{sa-reg-ffchain-h},
\textit{w-reg} and \textit{sa-reg-h} have the smallest fault magnitude and also have a small
contribution to the total number of \useacrplur{ff}, thus the gain in applying \useacr{seu}
protection is also small. An exception are the horizontal \useacr{sa} registers since they do show a
large sensitivity due to faults propagating to multiple columns.

Finally, the \bit{32} registers \textit{sa-reg-v}, \textit{sa-reg-ffchain-v} and \textit{accum-reg}
show the highest average fault magnitude. Additionally, these groups also include the highest number
of \useacrplur{ff} relative to the total accelerator. As such, these are interesting for
\useacr{seu} hardening, even though their fault propagation probability is quite low.

\section{Conclusion}
\label{sec:concl}

In this research, we performed a fault injection campaign on a \useacr{sa} based \useacr{dnn}
accelerator, with constrained inputs to mimic the distribution of the data observed in \useacr{dnn}
models. We concluded that the \bit{32} register groups have the highest fault magnitude as well as a
large contribution to the total amount of flip-flops. Given Amdahl's law, this means that these
registers present the largest gain for hardening against \useacrplur{seu}. Secondly, we also
observed that the usage of \useacr{relu} significantly impacts the fault propagation, making its
prevalent use in \useacrplur{dnn} a positive in terms of \useacr{seu} hardening.

Our analysis ensured that the accelerator is fully utilised, allowing for conclusions independent of
a \useacr{dnn}'s model architecture. However, typical \useacr{dnn}s might alter the utilisation,
leading to the processing chain to wait for the \useacr{sa} to finish, hereby skewing the fault
analysis. Therefore, future work will complement our analysis with real \useacr{dnn} workloads.

\bibliographystyle{IEEEtran}
\bibliography{references}
%



\end{document}